\documentclass{aa}  
\usepackage[utf8]{inputenc}

\usepackage{svg}
\usepackage{graphicx}
\usepackage{listings}
\usepackage{placeins}

\usepackage{xcolor}
\usepackage{xspace}
\usepackage{txfonts}
\usepackage{ulem}

\usepackage{hyperref}
\hypersetup{
    colorlinks=true,
    linkcolor=blue,
    filecolor=magenta,      
    urlcolor=cyan,
    pdftitle={Human vs. machine - 1:3. Joint analysis of classical and ML-based summary statistics of the Lyman-α Forest},
    pdfpagemode=FullScreen,
    }
\usepackage[nameinlink]{cleveref}

\bibpunct{(}{)}{;}{a}{}{,}

\newcommand{\Lya}{Ly\ensuremath{\alpha}F}
\newcommand{\E}[1]{\left\langle #1 \right\rangle}
\newcommand{\lyanna}{\textsc{Ly$\alpha$NNA}\xspace}
\newcommand{\B}[1]{\mathbf{#1}}

\begin{document}

   \title{Human vs. machine - 1:3. Joint analysis of classical and ML-based summary statistics of the Lyman-$\alpha$ forest}

   \author{S. Chang\inst{1,2}\thanks{\email{sookyung.chang@physik.lmu.de}}
   \and P. Nayak\inst{1,2}
   \and M. Walther\inst{1,2} 
   \and D. Gruen\inst{1,2,3}
          }

   \institute{University Observatory, Faculty of Physics, Ludwig-Maximilians-Universität, Scheinerstr. 1, 81679 Munich, Germany
   \and Munich Center for Machine Learning, Oettingenstr. 67, 80528 Munich, Germany
   \and Excellence Cluster ORIGINS, Boltzmannstr. 2, 85748 Garching, Germany
             }

   \date{Received , 2025; accepted}

  \abstract
   { In order to compress and more easily interpret Lyman-$\alpha$ forest (\Lya{}) datasets, summary statistics, e.g. the power spectrum, are commonly used. However, such summaries unavoidably lose some information, weakening the constraining power on parameters of interest. Recently, machine learning (ML)-based summary approaches have been proposed as an alternative to human-defined statistical measures. This raises a question: can ML-based summaries contain the full information captured by traditional statistics, and vice versa?
   
   In this study, we apply three human-defined techniques and one ML-based approach to summarize mock {\Lya} data from hydrodynamical simulations and infer two thermal parameters of the intergalactic medium, assuming a power-law temperature-density relation. We introduce a metric for measuring the improvement in the figure of merit when combining two summaries. 
   
   Consequently, we demonstrate that the ML-based summary approach not only contains almost all of the information from the human-defined statistics, but also that it provides significantly stronger constraints by a ratio of better than 1:3 in terms of the posterior volume on the temperature-density relation parameters.
   }

   \titlerunning{Classical and ML-based summary statistics of the Lyman-$\alpha$ forest}
   \authorrunning{Chang et al.}
   \maketitle

\section{Introduction}

\begin{figure*}
\centering
\includegraphics[width=\hsize]{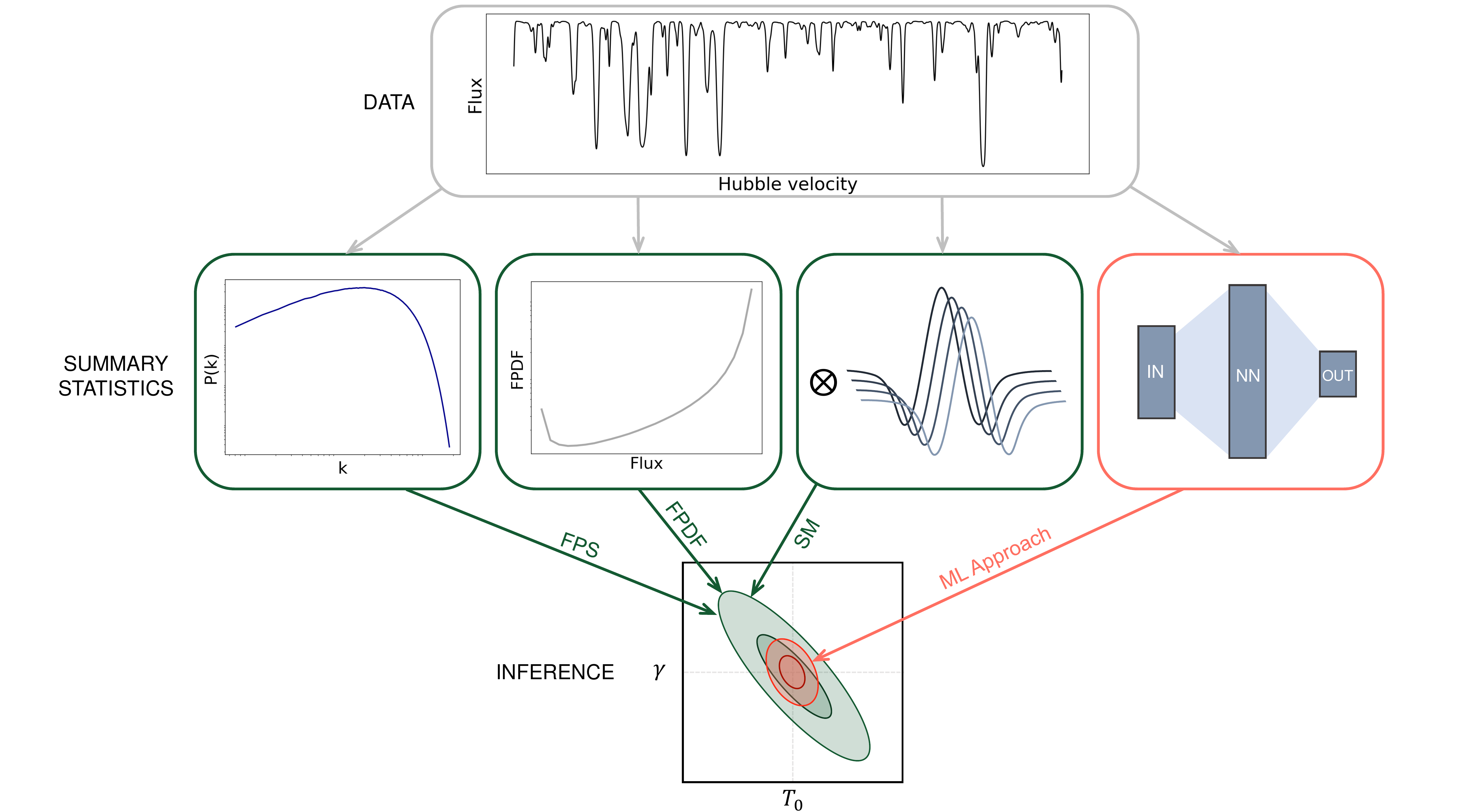}
  \caption{Human-defined statistics vs. ML-based approaches. This figure provides an overview of the workflow. The top panel represents our \Lya{} data, generated from a hydrodynamical simulation with periodic boundary conditions. We use four different summary statistics: the Flux Power Spectrum (FPS), the Flux Probability Density Function (FPDF), Scattering Moments (SM) of the flux, and an ML-based statistical method that compresses the full spectral information into summary vectors optimized for parameter inference. The joint posterior distributions from FPS, FPDF, and SM are compared to the posterior from an ML-based approach to assess whether it captures most of the information extracted by the human-defined summary statistics.}
    \label{fig:workflow}
\end{figure*}

The Lyman-$\alpha$ forest (Ly$\alpha$F, \citealt{10.1086/180695}) is a sequence of absorption features observed in the spectra of high-redshift quasars. As their light passes through neutral gas clouds of the intergalactic medium (IGM), the spectrum is redshifted and partially absorbed at the Ly$\alpha$ transition with a rest-frame wavelength of $\lambda_r = 1216\AA$. The variations in the density field along the quasar's line of sight lead to corresponding fluctuations in the absorption. Since the \Lya{} arises from these fluctuations in the IGM, its observations serve as a powerful probe of the cosmic gas distribution, thus enabling the inference of cosmological parameters (see \citealt{1998ARA&A..36..267R, 2025arXiv250314739D}).

The detailed structure of the absorption lines in the \Lya{} is significantly influenced by the intrinsic properties of the gas, such as its density distribution and thermal structure \citep{Hui_Equation_1997, 2015MNRAS.450.4081P, 2016MNRAS.456...47M}. With the rapid growth in volume and precision of the \Lya{} observations, such as those from eBOSS and DESI \citep{2016AJ....151...44D, 10.3847/1538-3881/aa7567, 10.3847/1538-3881/ac882b}, there has also been a resurgence of interest in reconstructing the thermal history of the IGM by using the \Lya{}. The \Lya{} serves as a valuable tool for measuring the IGM temperature, as the widths of the absorption lines are largely influenced by thermal effects such as Doppler broadening, peculiar velocities, and Jeans smoothing \citep{10.1146/annurev-astro-082214-122355, Kulkarni2015}. 
A wide variety of methods have been developed to probe the IGM thermal state with the use of the temperature-density relation (TDR) model \citep{Hui_Equation_1997}: the \Lya{} flux power spectrum (FPS, \citealt{1998ApJ...495...44C, 10.1086/317079, 2018ApJ...852...22W, 10.3847/1538-4357/aafad1, QMLE_DESI_DR1, FFT_DESI_DR1}), the flux probability density function (FPDF, \citealt{ 10.1086/170252, 10.1086/317079, 10.1111/j.1365-2966.2008.13114.x, Lee_2015, 10.1093/mnras/stw2917}), PDF of wavelet amplitudes \citep{10.1046/j.1365-8711.2002.05316.x, 2010ApJ...718..199L, 10.1111/j.1365-2966.2012.21223.x, 10.1093/mnras/stab2920}, the curvature statistics \citep{2011MNRAS.410.1096B, 10.1093/mnras/stu660}, and analyses based on decomposing the forest into individual Voigt profiles \citep{10.1088/2041-8205/757/2/L30, 2018ApJ...865...42H}. Since all of them are sensitive enough to changes in the widths of the \Lya{} absorption lines, these summary statistics are effective tools for capturing the information about the IGM thermal state. 
Using these human-defined summary statistics enables the measurement of the targeted properties of data while suppressing sensitivity to irrelevant features. In other words, selecting an appropriate statistic requires considering the relevant features of the parameters of interest. 
For the bulk of IGM gas, the TDR is expected to follow a tight power-law relation, typically parametrized as \citep{Hui_Equation_1997}: \begin{equation} T(\Delta)=T_0\Delta^{\gamma-1} \; , \label{eqn:tdr} \end{equation} with the overdensity $\Delta=\rho/\bar{\rho}$, the temperature at mean density $T_0$ and a logarithmic slope $\gamma-1$. A feature directly impacted by those parameters is the thermal broadening effect of the absorption lines in the \Lya{}. However, the summaries' sensitivity is not only determined by the Doppler broadening. Other IGM features, e.g. pressure smoothing of the gas which depends on its full thermal history, and background cosmological parameters can also influence the statistics. This implies that these statistics may capture only partial information sensitive to the parameters under study and potentially give rise to parameter degeneracies.

In contrast, the rise of ML technologies (see \citealt{Moriwaki_Machine_2023} for a recent review) has motivated the development of customized summary statistics derived from informative data for field-level inference \citep{10.1093/mnras/stac1786, Nayak_lyanna_2024, 2024A&A...690A.154M, 10.1093/mnras/stae2153}. In other applications of cosmological inference, ML-based summaries have been shown to greatly increase the constraining power and break parameter degeneracies \citep[e.g.][]{Gupta2018,Kacprzak2022}. This raises the question of whether neural networks trained on simulated \Lya{} data can similarly retain the most relevant features for parameter inference, and how their performance compares to classical methods. In this study, we demonstrate that an ML-based summary approach can capture almost all the information extracted by human-defined summary statistics. We employ three human-defined statistics, including the traditional techniques of FPS and FPDF along with the scattering transform derived from \citet{mallat_group_2012}, which more recently sparked interest in the cosmological community \citep{Cheng_newapproach_2020, tohfa_forecast_2023}. In addition, we employ the ML-based approach described in \citet{Nayak_lyanna_2024}. The authors trained a convolutional neural network on the \Lya{} data to infer the thermal parameters $T_0$ and $\gamma$ exclusively. We compare the three human-defined summaries\footnote{Note that a fourth summary, the curvature statistics, was tested in, but its information is already contained within the other human-defined summaries, see  \cref{app: curvature}.} to the ML-based summary by quantifying their information content based on identical test datasets. \Cref{fig:workflow} provides a diagram illustrating our workflow. We use simulated \Lya{} data as mock observations to perform inference using the three human-defined statistics and the ML-based statistic. We then compare their individual and joint posterior distributions. 

Studies on the relationships between various statistics for the \Lya{} in the context of IGM astrophysics have received limited attention. For example, combinations of various summary statistics have been explored in this area (see \citealt{gaikwad_consistent_2021}), but these comparisons typically disregard the full covariance structure and correlations among statistics. To address this gap, we measure the additional information when two statistics are combined and quantify their complementary information.

\section{Simulation Data}

The cosmological interpretation of the detailed structure of \Lya{} data heavily relies on hydrodynamical simulations to accurately model how the IGM properties evolve as cosmic structures form. In this work, we use outputs from the Nyx hydrodynamical simulation code for our thermal models \citep{Nyx_2013, 10.1093/mnras/stu2377}. The simulation box, at redshift $z = 2.2$, has a side length of 120 Mpc (comoving) and consists of $4096^3$ volumetric cells and dark matter particles. Cosmological parameters are fixed to $h = 0.7035$, $\omega_\mathrm{m} = \Omega_\mathrm{m} h^2 = 0.1589$, $\omega_\mathrm{b} = \Omega_\mathrm{b} h^2 = 0.0223$, $10^{9}A_\mathrm{s} = 1.4258$, $n_\mathrm{s} = 1.0327$, and $\lambda_P(z=2.2)=63.7 \text{ kpc}$. The simulation box is the same as that used by \citet{Nayak_lyanna_2024}; the simulation suite is described in \cite{arXiv:2412.05372}. 

Our mock dataset comprises line-of-sight spectra generated with fixed fiducial thermal parameters. The optical depth $\tau$ values are rescaled by a constant factor so that the mean \Lya{} transmission in the full set of skewers matches its observed value by \cite{2013MNRAS.430.2067B}. We rescaled the temperatures inside the simulation box with a density-dependent function according to the procedure described in \citet{Nayak_lyanna_2024} to generate a regular grid of thermal models with different TDRs.

To mimic observational limitations and minimize the impact of numerical noise in the simulated data, modes larger than $k_{\text{max}}=0.182\,\mathrm{s/km}$ are removed from the spectra and the spectra are re-binned by performing 8-pixel averages. 

\section{Summary Statistics}

Summarizing data plays a crucial role in extracting meaningful patterns from complex observations. These summary statistics offer informative representations that emphasize specific physical features of the data. In this study, we employ four different statistics to summarize the \Lya{} data---FPS, FPDF, SM, and an ML-based approach---in order to emphasize structural characteristics shaped by the thermal parameters $T_0$ and $\gamma$ and downplay irrelevant features. 

\subsection{FPS}
We define FPS as the variance of the Fourier-transformed flux contrast, $P_F(k) \propto \E{|\tilde{\delta}_F(k)|^2}$ for a given wavenumber $k$, between different lines-of-sight.
Here, $\delta_F(v)$ is expressed as the contrast in the transmitted flux at Hubble velocity $v$ along a line-of-sight, $\delta_F(v) = (F(v)-\E{F}_v)/\E{F}_v$, and $\tilde{\delta}_F(k)$ represents the Fourier transform of $\delta_F(v)$. We use $P_{F,i}(k) \sim |\tilde{\delta}^i_{F,k}|^2$ as the FPS summary statistic for individual lines-of-sight, covering wavenumbers from the fundamental mode at $k \sim 0.0007$ s/km to the resolution cut at $k \sim 0.1822$ s/km. Each $P_{F,i}(k)$ then has a length of 256.

\subsection{FPDF}
We compute the FPDF statistic as the histogram of the transmitted flux with 25 equal-width bins from 0 to 1. The number of bins is selected given that using a larger number of bins requires more samples for the posterior distribution to converge. With a total of $\sim 10^{5}$ spectra, the FPDF converges sufficiently when using 25 bins. We omit the last bin, as it is fully degenerate with the others due to the normalization property of probability distribution functions, which is normalized to an integral of 1. The bins are slightly narrower than previous measurements, e.g. by \citet{Lee_2015}.

\subsection{Scattering Moments}
We compute the first-order scattering moments by averaging the output of the first wavelet transform of $\delta_F(v)$ over velocity, $SX(j_1) =\E{|\delta_F \ast \psi_{j_1}|}$. Here, the first set of wavelet filters is denoted by $\psi_{j_1}$. The second-order scattering moments can partially recover and preserve information from the first wavelet transform, and are defined as $SX(j_1, j_2)=\E{||\delta_F \ast \psi_{j_1}|\ast \psi_{j_2}|}$. In this work, the same set of 9 wavelet filters is used for both $\psi_{j_1}$ and $\psi_{j_2}$. For a detailed calculation, see \Cref{app: scattering moments}. Henceforth, the term SM1 will denote the statistic comprising all first-order scattering moments, $SX(j_1)$ for $j_1 = 0, \dots, 8$. The term SM2 will denote the statistic constructed from $SX(j_1, j_2)/SX(j_1)$, where $0 \leq j_1 < j_2 \leq 8$.

\subsection{ML-based approach}
To generate summary vectors using ML, we apply the method proposed by \citet{Nayak_lyanna_2024}. The authors trained a convolutional neural network (CNN) on hydrodynamical \Lya{} simulation data labeled with the TDR parameters $T_0$ and $\gamma$. They trained the CNN to recognize patterns that vary with $T_0$ and $\gamma$ so that the output may contain information about the parameters. The architecture of the CNN consists of four residual blocks and a total of 136,784 trainable parameters, with leaky ReLU used as the activation function. The input size is 512 (the length of the simulated spectrum), and the output size is 5, representing a direct estimation of the parameters and a parameter covariance matrix. In this study, we use only the first two outputs for summary vectors because they represent a direct estimation of $T_0$ and $\gamma$. Henceforth, this ML-based summary statistic will be referred to as \lyanna, following the same convention established by \citet{Nayak_lyanna_2024}.

\section{Posterior Analysis}
\label{sec: posterior analysis}
In order to constrain the thermal parameters $T_0$ and $\gamma$, we employ Bayesian inference. The posterior distribution is defined as
\begin{equation}
    Pr(\B{\Theta}|\B{S})= \frac{Pr(\B{S}|\B{\Theta}) Pr(\B{\Theta})}{Pr(\B{S})},
    \label{equ: Bayesian inference formula}
\end{equation}
where $\B{\Theta} = ( T_0, \gamma )$ denotes the parameter vector, and $\B{S}$ represents the observed summary statistic, such as the FPS. The prior distribution $Pr(\B{\Theta})$ is assumed to be flat over the ranges $T_0 \in [6000\text{K}, 15000\text{K}]$ and $\gamma \in [1.30, 1.66]$, covering the extent of our simulation grid. The likelihood $Pr(\B{S}|\B{\Theta})$ is modeled by assuming a multivariate Gaussian distribution for $\B{S}$. The log-likelihood is then defined as
\begin{equation}
\label{equ: likelihood}
    \log\mathcal{L}=\log Pr(\B{S}|\B{\Theta})= \Lambda -\frac{1}{2} \Delta_{S}^T \B{\Sigma}^{-1}\Delta_{S},
\end{equation}
where 
\begin{equation}
    \Delta_{S}=(\delta_0,\cdots, \delta_n).
    \label{equ: difference}
\end{equation}
$\delta_n=\E{s}_i -\E{s_{\text{mock}}}_i$ represents the deviation of the n-th component between the averaged summary vectors from the mock data and the thermal models. $\B{\Sigma}$ is the covariance matrix of the $\B{S}_{\text{mock}}$ rescaled with the uncertainty range corresponding to a 1$\sigma$ equivalent of 100 spectra. We apply cubic interpolation\footnote{For more details on the cubic interpolation, refer to \texttt{scipy.interpolate.RectBivariateSpline}} 
to $\B{S}$ in parameter space in order to estimate its values between grid points, and we sample the posterior distribution using affine-invariant Markov Chain Monte Carlo as implemented in \texttt{emcee} \citep{Foreman_emcee_2013}.
In order to combine information from multiple summary statistics, a joint likelihood $\mathcal{L}_\mathrm{joint}$ is defined as
\begin{equation}
    \log\mathcal{L}_\mathrm{joint} \propto -\frac{1}{2}\left(\Delta^T_{S_1+\dots+S_n} \B{\Sigma}_{S_1+\dots+S_n}^{-1}\Delta_{S_1+\dots+S_n}\right),
    \label{equ: joint likelihood}
\end{equation}
where
\begin{equation*}
    \Delta_{S_1+\dots+S_n} = (\Delta_{S_1},\dots, \Delta_{S_n}).
\end{equation*}
$\Delta_{S_1+\dots+S_n}$ expresses a concatenation of individual $\Delta_S$ (Equ.~\Ref{equ: difference}). We estimate the covariance matrix $\mathbf{\Sigma}_{S_1+\dots+S_n}$ from the concatenated vector $(\B{S}_1,\dots,\B{S}_n)$.
This process combines information from $n$ different posterior distributions into a single joint posterior distribution. In this study, we use the joint covariance matrices for the multiple combinations of the summary statistics FPS, FPDF, SM1, and SM2.
One of the corresponding correlation matrices $\mathbf{\Sigma}_{\text{FPS+FPDF+SM1+SM2}}$ is shown in \Cref{app:corrmat}.

Note that $\mathbf{\Sigma}$ represents the covariance matrix of the summaries, not the parameter covariance matrix $\mathbf{C}$, which is estimated via MCMC under the assumption of Gaussianity. The determinant $|\mathbf{C}|$, known as the \textit{generalized variance} \citep{wilks_certain_1932}, provides a scalar measure of the uncertainty in the inferred parameters. It offers a geometric interpretation of ``spread'' in higher-dimensional spaces. In this study, we adopt $1/\sqrt{|\mathbf{C}|}$ as a figure of merit (FoM) to quantify and compare the informational content of different summary statistics.

\begin{figure}
    \centering
    \includegraphics[width=1.0\columnwidth]{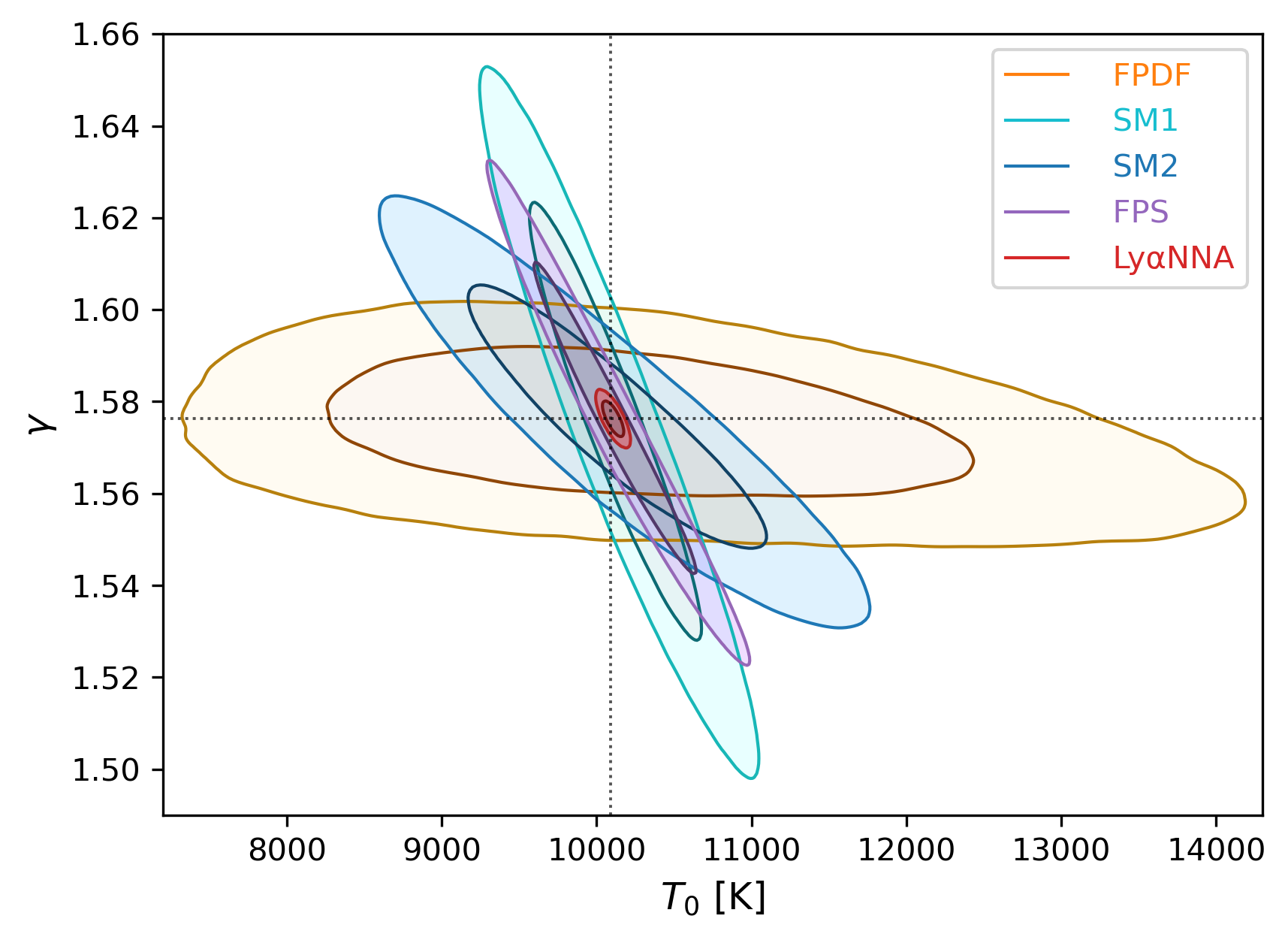}
    \caption{Posterior distributions of FPDF, SM1, SM2, FPS, and \lyanna (ML-based). Compared to other statistics, FPDF exhibits minimal dependence on the parameter $T_0$. FPS shows a degeneracy orientation more similar to SM1 than to SM2. The \lyanna (ML-based) posterior provides a significantly stronger constraint on the TDR parameters compared to other statistics.}
    \label{fig: FPS_FPDF_cur_sansa}
\end{figure}

\begin{figure}
    \centering
    \includegraphics[width=1.0\columnwidth]{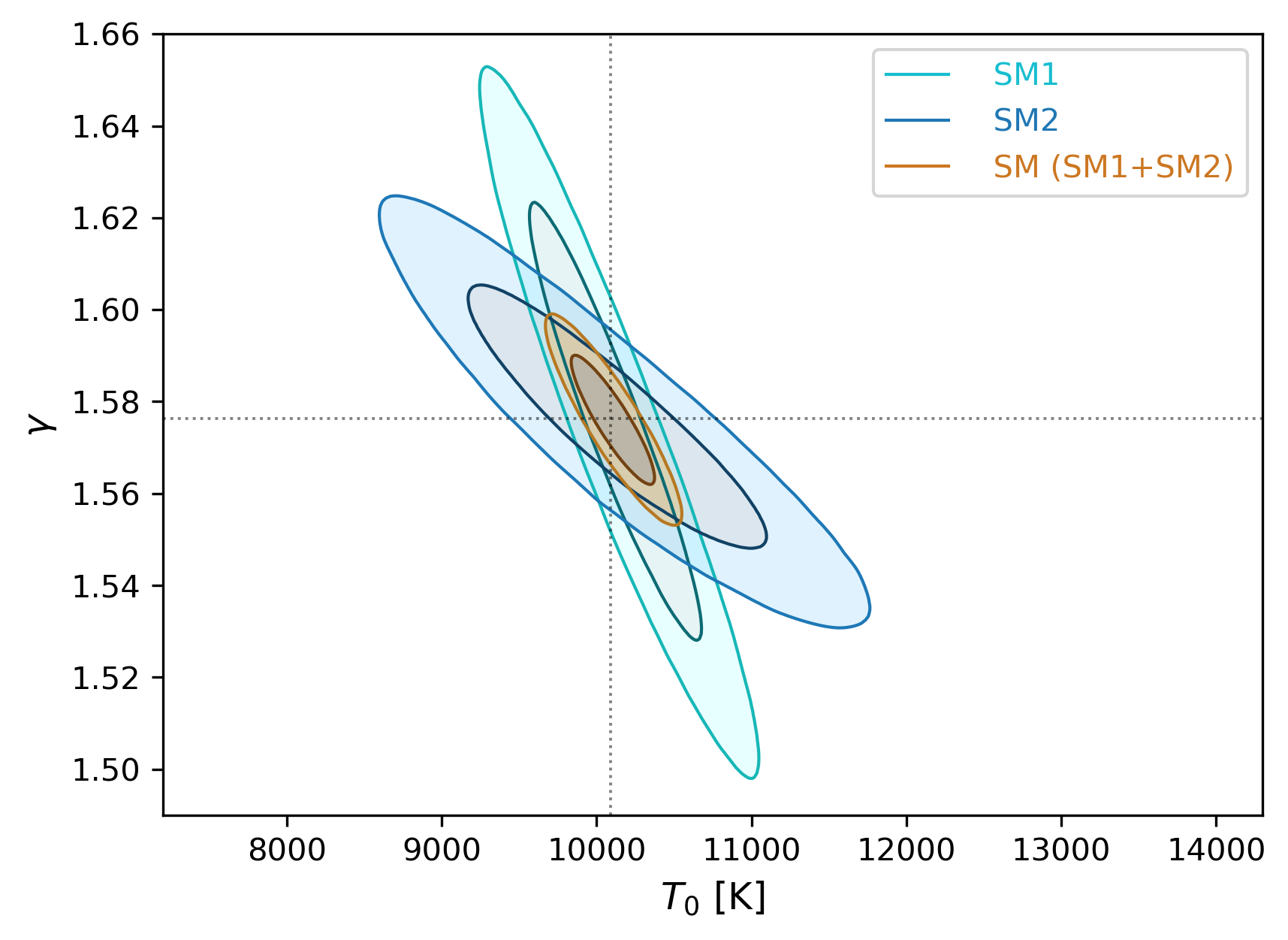}
    \caption{Individual posterior distributions of SM1 and SM2, together with their joint posterior distribution. These distributions show that SM1 is more sensitive to $T_0$ than SM2, while SM2 is more influenced by $\gamma$ than SM1. The different orientations in their posterior degeneracy indicate that SM1 and SM2 provide complementary information. The joint posterior of SM1 and SM2 shrinks significantly compared to the individual posteriors. This strong constraint on the TDR parameters suggests that SM1 and SM2 contain largely independent information.}
    \label{fig: joint sm1 and sm2}
\end{figure}

\section{Results: Posterior Distribution Comparisons}
\label{sec:results}

In \Cref{fig: FPS_FPDF_cur_sansa}, we present the posterior distributions of the following statistics individually: FPS, FPDF, SM1, SM2, and \lyanna. The variation in the orientation of their posterior degeneracy suggests that different summary statistics identify distinct structures and patterns in the flux related to the parameters $\gamma$ and $T_0$ (cf.~\autoref{eqn:tdr}). The degeneracy of FPDF lies predominantly along the $T_0$ axis, implying weaker sensitivity to $T_0$ compared to other summary statistics. Among the statistics, \lyanna stands out with a significantly tighter constraint on both $T_0$ and $\gamma$. Thus, \lyanna satisfies a necessary condition for encompassing all information from the other statistics: its information content is equal to or greater than that of any other summaries.

\Cref{fig: joint sm1 and sm2} shows the joint posterior distribution for SM1 and SM2, along with their individual distributions. Note that SM2 was calculated based on the output of the first scattering transform (see \Cref{app: scattering moments}). To quantify the complementary information between SM1 and SM2, we utilize their joint posterior distribution, which encapsulates the combined information provided by both of them. Here, there are two extreme scenarios:
\begin{itemize}
    \item The joint posterior of SM1 and SM2 is comparable in volume to that of either alone.
    \item The joint posterior of SM1 and SM2 is significantly smaller in volume than that of either alone.
\end{itemize}
For the first case, since the joint posteriors show similar performance in constraining parameters to that of either SM1 or SM2 alone, this suggests little complementary information; when SM2(SM1) is combined with SM1(SM2), SM2(SM1) does add little information to SM1(SM2). On the other hand, in the second case, since the volumes of the joint posteriors shrink significantly, it indicates that SM2(SM1) contains substantial independent information that SM1(SM2) does not capture, leading to a large amount of complementary information. \Cref{fig: joint sm1 and sm2} displays a noticeable difference between the joint posterior and each individual posterior distribution, indicating that SM2 provides complementary information to SM1. The joint posterior distribution of SM1 and SM2 will be referred to as the scattering moments (SM) posterior distribution when we compare its performance to that of other summary statistics. 

\subsection{Comparison with \texorpdfstring{\lyanna}{lyanna}}

\begin{figure}
    \centering
    \includegraphics[width=\hsize]{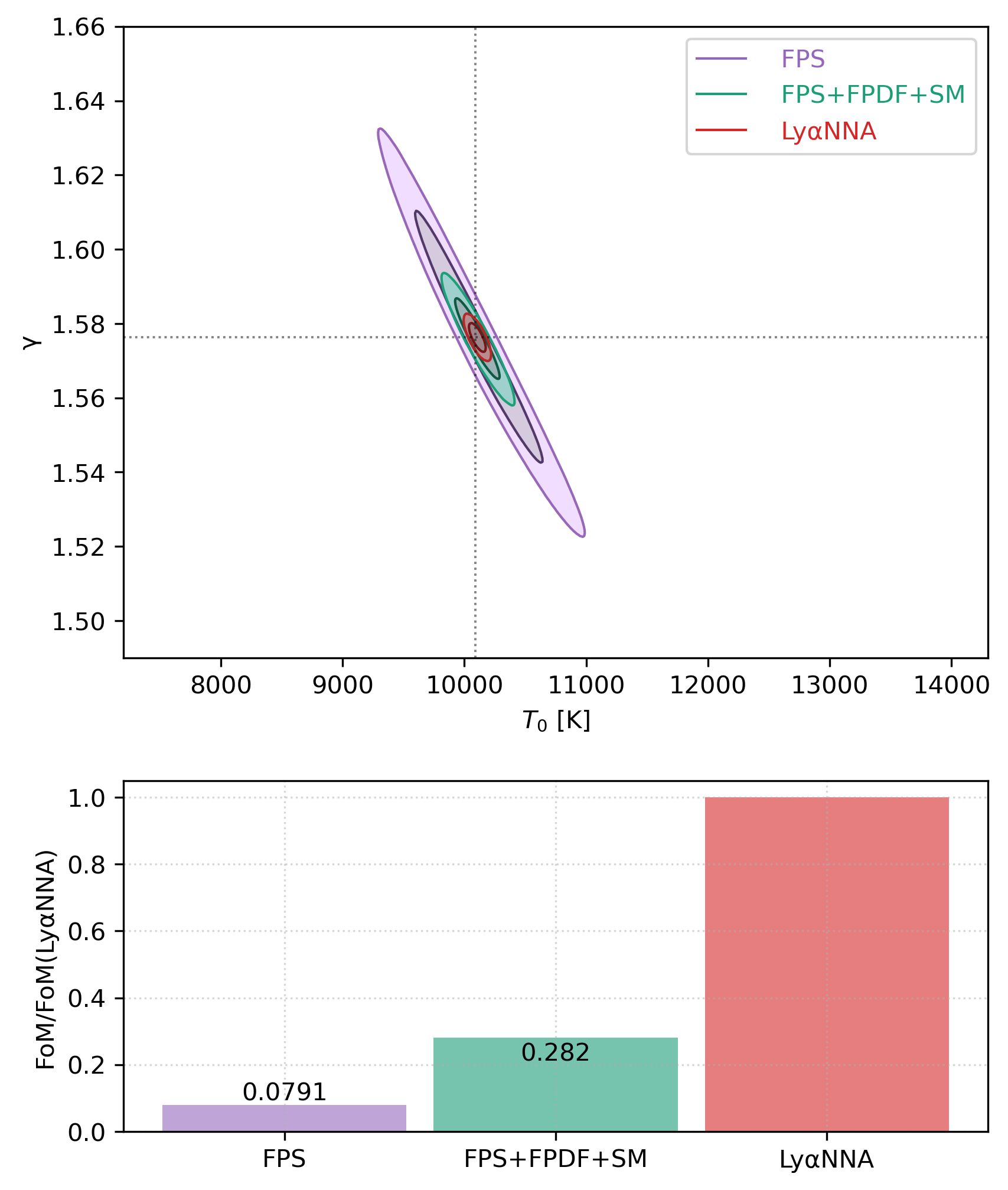}
    \caption{Posterior distributions inferred from FPS and \lyanna (ML-based) statistics, as well as the joint posterior from combining FPS, FPDF, and SM (top panel). The bottom panel displays the corresponding FoMs derived from these posteriors. \lyanna (ML-based) captures significantly more information than the combination of FPS, FPDF, and SM.}
    \label{fig: sansa_bar_graph}
\end{figure}

\begin{figure}
    \centering
    \includegraphics[width=\hsize]{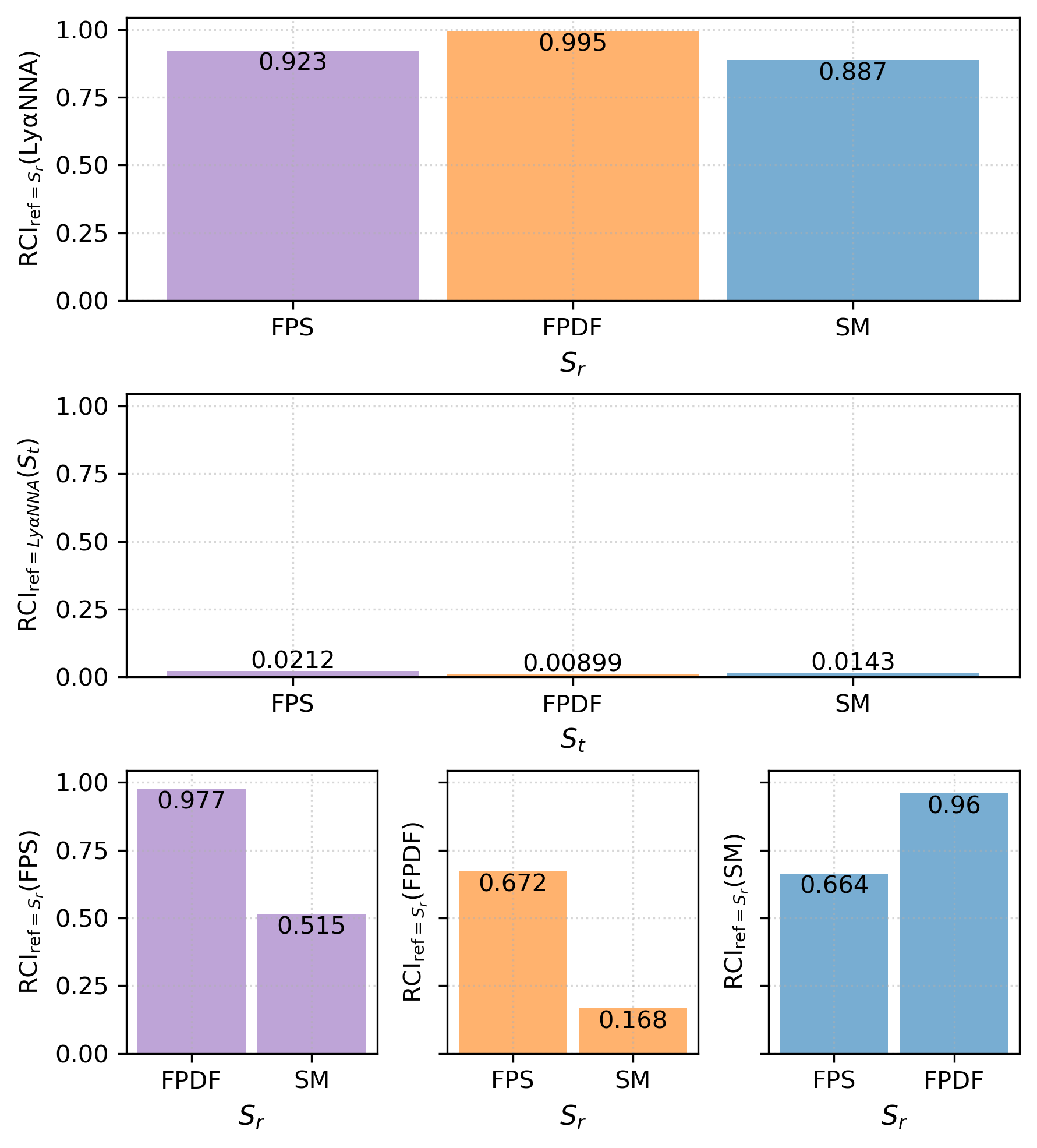}
    \caption{Relative complementarity indices $\text{RCI}_{\text{ref}=S_r}(S_t)$, among FPS, FPDF, SM, and \lyanna (ML-based). The top panel displays $\text{RCI}_{\text{ref}=S_r}(\lyanna)$; the amount of complementary information when \lyanna (ML-based) is added to a target statistic, in this case FPS, FPDF, or SM. On the other hand, the middle panel shows $\text{RCI}_{\text{ref}=\lyanna}(S_t)$; the amount of complementary information when a reference statistic is added to \lyanna (ML-based). The bottom panels display the relative complementarity index among FPS, FPDF, and SM.}
    \label{fig:rci}
\end{figure}

In the top panel of \Cref{fig: sansa_bar_graph}, we show the posterior distributions of FPS and \lyanna, together with the joint posterior distribution of FPS, FPDF, and SM. The posterior volume of FPS on the parameters decreases when combined with FPDF and SM, suggesting that FPDF and SM provide substantial independent information beyond what is captured by FPS alone. This decrease can be further quantified by comparing the FoMs of each summary and their combination (bottom panel of \Cref{fig: sansa_bar_graph}, normalized to the FoM of \lyanna). Relative to \lyanna, the FPS reaches about 10\% of the FoM, while the combination of FPS, FPDF, and SM reaches almost 30\%. The FoM of \lyanna is significantly larger than that of FPS+FPDF+SM, indicating that \lyanna provides substantially more information related to $T_0$ and $\gamma$ than the combination of three human-defined summary statistics considered in this work, i.e. FPS, FPDF, and SM.

To further examine whether \lyanna captures all information contained in these human-defined statistics, we introduce a metric to quantify the additional information when two statistics are combined. This metric, referred to as \textit{relative complementarity index}, is defined by

\begin{equation}
    \text{RCI}_{\text{ref}=S_r}(S_t) = \frac{\text{FoM}(S_r+S_t)-\text{FoM}(S_r)}{\text{FoM}(S_r+S_t)} = 1-\frac{\sqrt{|\B{C}_{S_r+S_t}|}}{\sqrt{|\B{C}_{S_r}|}},
    \label{equ:rci}
\end{equation}

Here, $S_r$ and $S_t$ represent the reference and target summary statistics, respectively. $\sqrt{|\B{C}_{S_r+S_t}|}$ quantifies the volume of the joint posterior derived from both $S_r$ and $S_t$. The relative complementarity index measures how much the posterior volume of $S_r$ is reduced when it is combined with  $S_t$. For instance, if $S_t$ adds little information to $S_r$, the relative complementarity index approaches 0 as the posterior volume of the combination of $S_r$ and $S_t$ converges to that of $S_r$. In contrast, if $S_t$ adds a significant amount of information to $S_r$, the relative complementarity index approaches 1 as the posterior volume of $S_r$ is much wider than the joint posterior volume of $S_r$ and $S_t$. Such a case is likely when $S_t$ contains substantial independent information. This tendency is clearly illustrated by $\text{RCI}_{\text{ref}=S_r}(\lyanna)$ in the top panel of \Cref{fig:rci}. Since \lyanna{} constrains the parameters $T_0$ and $\gamma$ very strongly, it likely adds a significant amount of information to the reference statistics FPS, FPDF, and SM as evidenced by the near-unity values of $\text{RCI}_{\text{ref}=S_r}(\lyanna)$. Moreover, since SM has a larger FoM than FPS and FPDF (see \Cref{app:alltheFoM}), $\text{RCI}_{\text{ref}=\text{SM}}(\lyanna)$ is lower than others.

The middle panel of \Cref{fig:rci} represents $\text{RCI}_{\text{ref}=\lyanna}(S_t)$, indicating the amount of information added to \lyanna{} when a human-defined summary $S_t$ is combined with it. Their values are close to zero because the human-defined summary statistics---FPS, FPDF, and SM---contribute only marginally to the information already contained in \lyanna. A notable observation is that SM exhibits a smaller value than FPS, despite its higher total information. This implies greater redundancy between SM and \lyanna than between FPS and \lyanna. In the bottom panels, $\text{RCI}_{\text{ref}=S_r}(\text{FPS})$, $\text{RCI}_{\text{ref}=S_r}(\text{FPDF})$, and $\text{RCI}_{\text{ref}=S_r}(\text{SM})$ demonstrate that reference statistics with a smaller individual FoM are associated with a greater relative complementarity index. Moreover, the greater value of $\text{RCI}_{\text{ref}=\text{FPDF}}(\text{FPS})$ relative to $\text{RCI}_{\text{ref}=\text{FPDF}}(\text{SM})$ implies that SM shares more redundant information with FPDF than FPS does, especially given that SM’s FoM is greater than that of FPS (see \Cref{app:alltheFoM}). Similarly, the values of $\text{RCI}_{\text{ref}=\text{FPS}}(\text{FPDF})$ and $\text{RCI}_{\text{ref}=\text{FPS}}(\text{SM})$ indicate that the information contained in SM is more redundant with FPS than the information contained in FPDF.

\section{Discussion and Conclusion}
We compared an ML-based approach ({\lyanna}, \citealt{Nayak_lyanna_2024}) with three human-defined summary statistics: FPS, FPDF, and SM. The main results are summarized below.
\begin{itemize}
    \item \lyanna has the strongest constraint compared to any of the tested human-defined approaches individually (\cref{fig: FPS_FPDF_cur_sansa}).
    \item \lyanna even contains more information than the total joint posterior of FPS, FPDF, and SM (\cref{fig: sansa_bar_graph}), by a factor of more than 3 in terms of FoM, i.e.~of the inverse volume of the posterior of the TDR parameters $T_0$ and $\gamma$ (\autoref{eqn:tdr}).
    \item The relative complementarity index confirms that the total information in each human-defined statistic is nearly fully encompassed by \lyanna; FPS, FPDF, and SM contain very little substantial independent information beyond what is already captured by \lyanna.
    \item There is a substantial overlap of information and substantial cross-correlation between the different human-defined summary statistics that needs to be accounted for when interpreting results from any combination of statistics.
\end{itemize}

We introduced a method to compare different statistics by quantifying the independent information they provide. This method demonstrates that ML-based approaches can contain most of the information extracted by human-defined statistics. We note that the quantitative findings are specific to the features of interest, which in this work are the two parameters of the power-law temperature-density relation. Also, the current results were based on simulated mock data without noise. Extending this analysis to include other parameters and realistically noisy data is left for future work. The promise of ML-based summary statistics demonstrated here strongly motivates such further study.

\begin{acknowledgements} 
We thank all the members of the chair of Astrophysics, Cosmology and Artificial Intelligence (ACAI) at LMU Munich for their continued support and very interesting discussions. 
We acknowledge the Faculty of Physics of LMU Munich for making computational resources available for this work. %
We acknowledge PRACE for awarding us access to Joliot-Curie at GENCI@CEA, France via proposal 2019204900. 
We also acknowledge support from the Excellence Cluster ORIGINS which is funded by the Deutsche Forschungsgemeinschaft (DFG, German Research Foundation) under Germany’s Excellence Strategy -- EXC-2094 -- 390783311. 
PN thanks the German Academic Exchange Service (DAAD) for providing a scholarship to carry out this research. 
MW acknowledges support by the project AIM@LMU funded by the German Federal Ministry of Education and Research (BMBF) under the grant number 16DHBKI013.

\end{acknowledgements} 

\section*{Data availability}
The simulation data set can be provided upon reasonable request.

\bibliographystyle{aa}
\bibliography{reference}

\begin{appendix}
\twocolumn

\section{Scattering Moments}
\label{app: scattering moments}
Scattering moments are derived from the scattering transform introduced by \citet{mallat_group_2012}, which iteratively applies two main operations: modulus and convolution with a family of wavelet functions. The method has shown reliable performance in different applications, such as audio classification. It preserves time-invariant features and recovers high-frequency information usually lost with conventional compression \citep{anden_deep_2014}. To capture irregular yet self-similar properties in time, \citet{bruna_intermittent_2015} introduced first- and second-order scattering moments by iteratively applying wavelet transforms and nonlinear modulus operations. \citet{Cheng_newapproach_2020} applied scattering moments to infer cosmological parameters in the context of weak lensing. Following this, \citet{tohfa_forecast_2023} demonstrated the effectiveness of the scattering transform in the analysis of \Lya{} data, achieving tighter constraints than FPS for four cosmological parameters.

In this work, we compute the scattering moments using the open-source library \texttt{Kymatio} (see \url{https://www.kymat.io/}). The steps for obtaining the first- and second-order moments are described in the following.
We define a wavelet function $\psi(v)$ that satisfies the conditions $\int \psi(v)dv = 0$ and $|\psi(v)| = \mathcal{O}\left((1 + |v|^2)^{-1}\right)$. Wavelets at different scales are constructed by scaling $\psi(v)$ by $2^j$, for integer values of $j$,
\begin{equation}
     \psi_j(v)\equiv 2^{-j}\psi\left(2^{-j}v\right).
\end{equation}
As $j$ increases, the wavelet $\psi(v)$ becomes broader in width and lower in amplitude. The first wavelet transform of the function $\delta_F(v)$ is then defined as
\begin{equation}
    \text{WT}^{1\text{st}}(j_1)=\left|\delta_F(v) \ast \psi_{j_1} (v)\right| = \left|\int dv^{'} \delta_F(v^{'}) \psi_{j_1}(v-v^{'})\right|.
    \label{equ: first wavelet transform}
\end{equation}
The first set of wavelet filters is denoted by $\psi_{j_1}$. The corresponding first-order scattering moments, $SX(j_1)$, are computed by averaging the output of the first wavelet transform over $v$,
\begin{equation}
    SX(j_1)=\E{|\delta_F \ast \psi_{j_1}|}.
    \label{equ: 1st-order scattering moments}
\end{equation}
As a result of averaging, first-order scattering moments lack information on irregular patterns or short-lived characteristics across spatial locations. Second-order scattering moments, however, can partially recover and preserve this information \citep{bruna_intermittent_2015}. The second-order scattering moments are defined as
\begin{equation}
    SX(j_1, j_2) = \E{\left||\delta_F \ast \psi_{j_1}|\ast \psi_{j_2}\right|}.
\end{equation}
Here, $\psi_{j_2}$ denotes the second set of wavelet filters, which, in this case, is identical to the first set. The total number of first- and second-order scattering moments depends on the number of filters used. For the second-order case, only configurations where $j_2 > j_1$ are considered, as $SX(j_1, j_2)$ rapidly approach zero when $j_2 < j_1$ and the difference $j_1 - j_2$ increases. The code for this computation can be found at \url{https://github.com/SookyungChang/summary-vs-ML-statistic}.

\section{Correlation Matrix}
\label{app:corrmat}
\begin{figure}
    \centering
    \includegraphics[width=0.95\columnwidth]{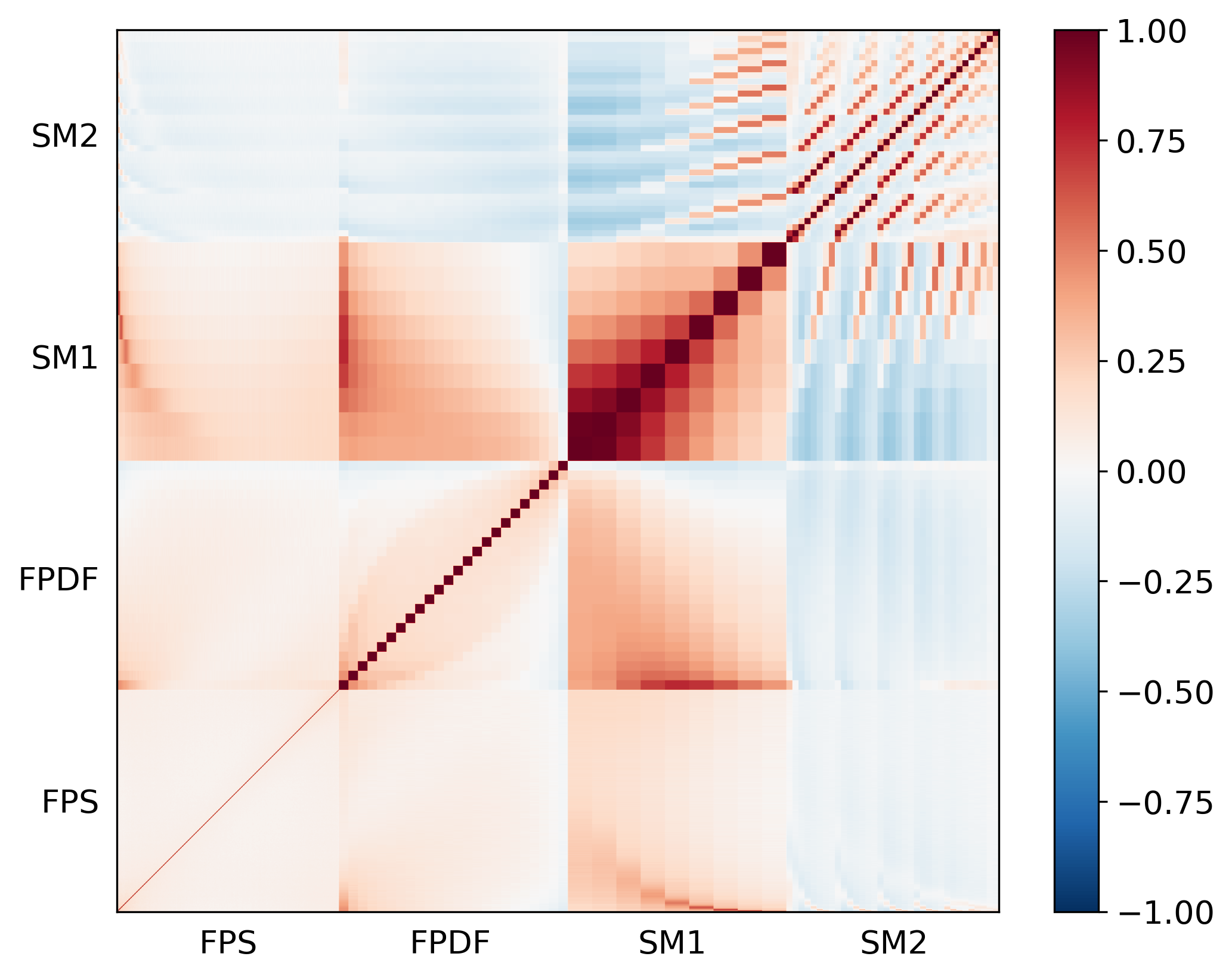}
    \caption{Correlation matrix derived from the joint correlation matrix of FPS, FPDF, SM1, and SM2. The diagonal panels show the correlation matrices of FPS, FPDF, SM1, and SM2 alone, arranged from bottom-left to top-right. The off-diagonal panels illustrate the correlation between pairs of summaries; for example, the top-left panel represents the correlation between SM2 and FPS. The lengths of the summary vectors are FPS: 256, FPDF: 24, SM1: 9, and SM2: 35.}
    \label{fig: correlation matrix}
\end{figure}
The covariance matrix $\mathbf{\Sigma}$ plays an important role in inference on Gaussian likelihoods, yet a considerable amount of research tends to disregard the cross-summary elements of the joint covariance matrix between different summaries of the \Lya{} (e.g. \citealt{gaikwad_consistent_2021}). \Cref{fig: correlation matrix} shows the correlation matrix derived from $\mathbf{\Sigma}_{\text{FPS}+\text{FPDF}+\text{SM1}+\text{SM2}}$. In the diagonal panels, the correlation matrices for FPS, FPDF, SM1, and SM2 alone are listed, while each off-diagonal panel presents the correlation between a pair of summaries. Here, there are six pairs: SM2 \& FPS, SM1 \& FPS, FPDF \& FPS, SM2 \& FPDF, SM1 \& FPDF, and SM2 \& SM1. Since their summary vectors vary in length, each panel is displayed at a different resolution. 

\section{FoM: Additional Summary Combinations}
\label{app:alltheFoM}

\begin{figure}
    \centering
    \includegraphics[width=\hsize]{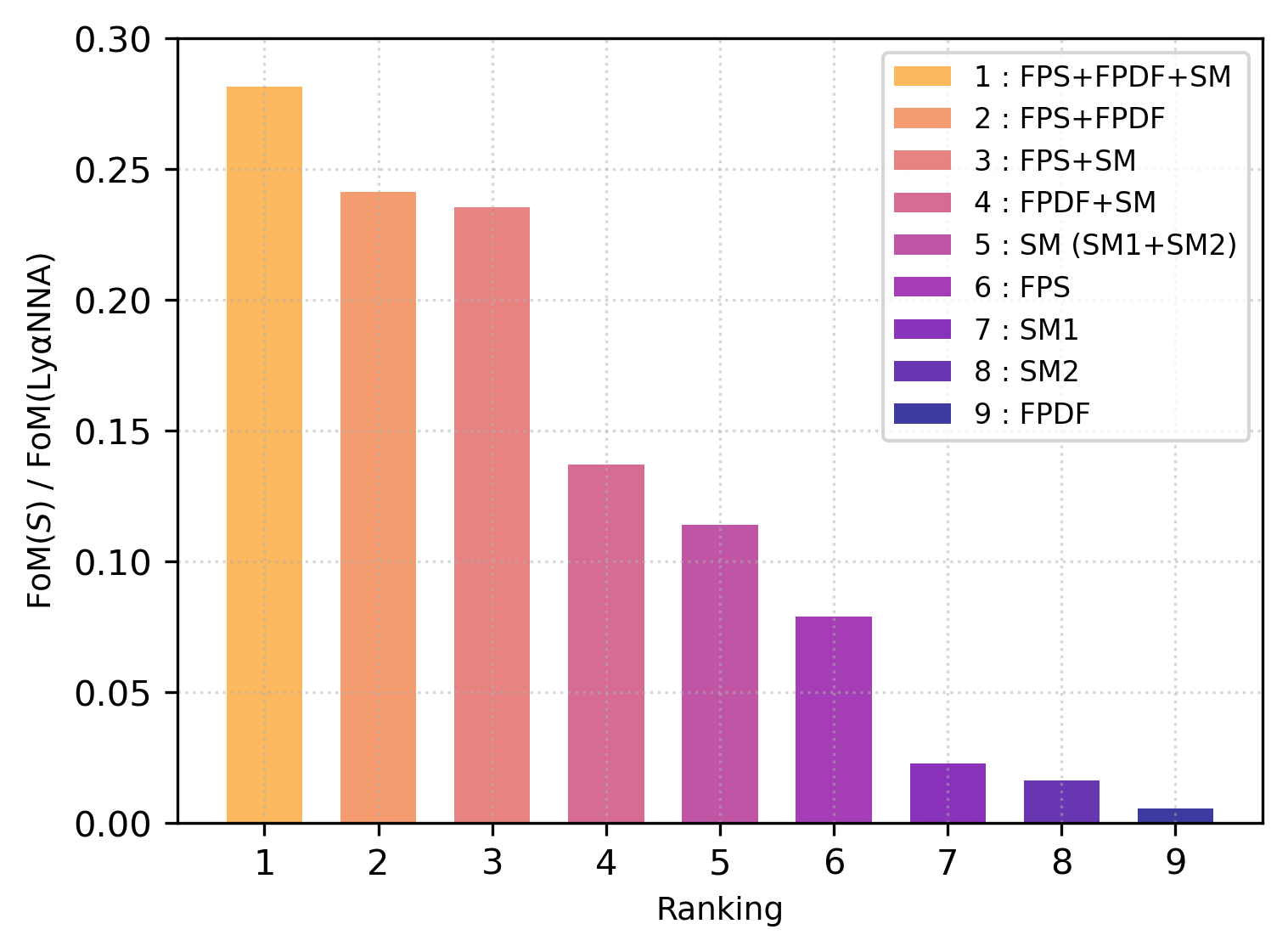}
    \caption{Ranking of the different combinations of summaries by FoM. All FoM values are normalized to the FoM of \lyanna.}
    \label{fig:all_FoM}
\end{figure}

\begin{figure}
    \centering
    \includegraphics[width=\hsize]{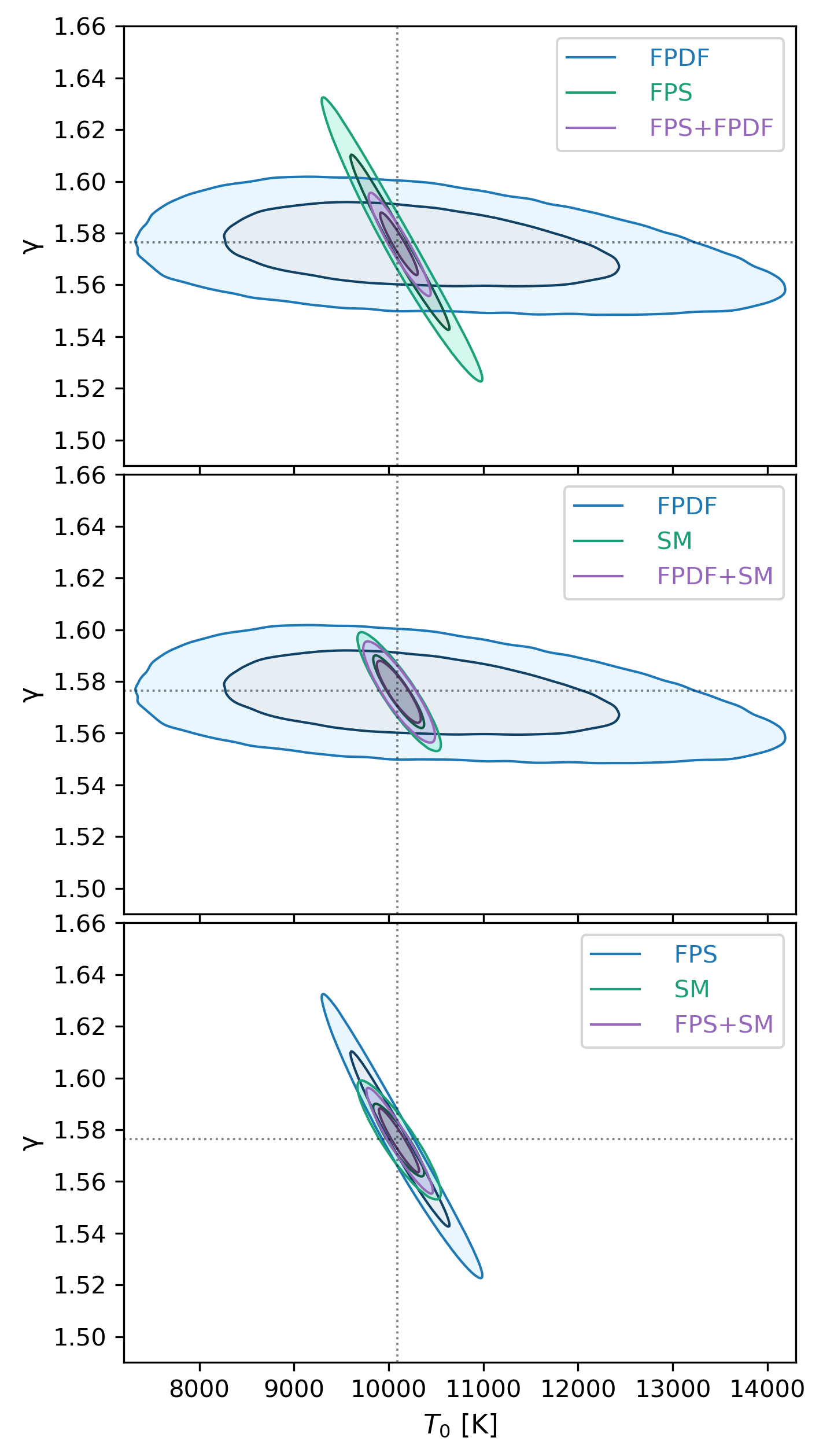}
    \caption{Joint posterior distributions of the combinations between FPS, FPDF, and SM with the respective individual posteriors.}
    \label{fig:fps_fpdf_sm}
\end{figure}

\Cref{fig:all_FoM} presents a FoM ranking for different combinations of summaries and for the individual FPS, FPDF, SM1, and SM2. Their FoMs are normalized by that of \lyanna, highlighting the increase in information when more summaries are combined. Among the individual statistics, the information content follows the order: FPS, SM1, SM2, and FPDF. For the cases FPS+FPDF+SM, SM, FPS, SM1, SM2, and FPDF, the actual posterior distributions are presented in \Cref{sec:results}. For the rest of the cases, the joint posterior distributions are displayed with their respective individual posteriors in \Cref{fig:fps_fpdf_sm}.

\section{SM and Curvature statistic}
\label{app: curvature}
\begin{figure}
    \centering
    \includegraphics[width=\hsize]{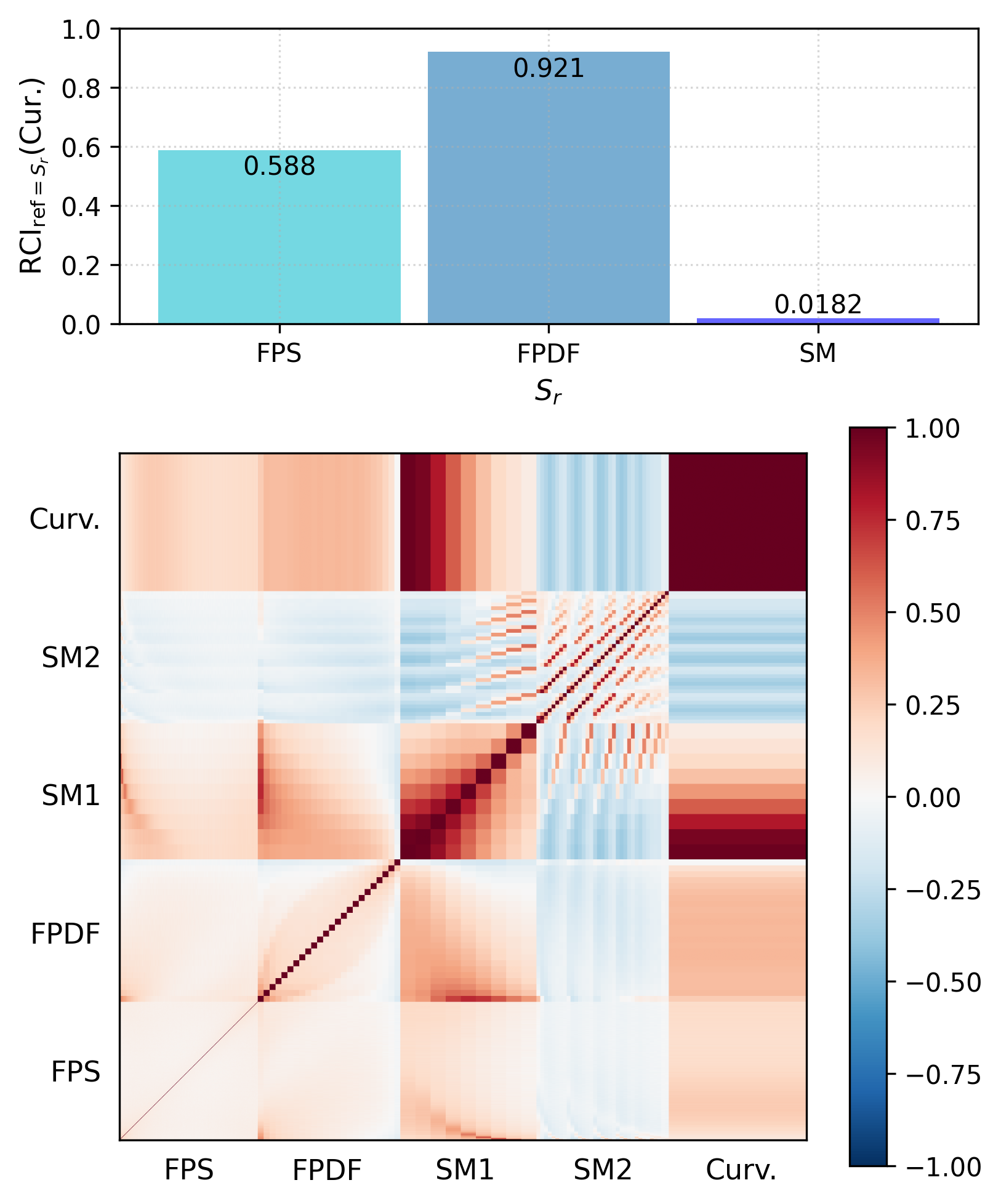}
    \caption{Relative complementarity index when curvature is the target statistic and the reference statistic is FPS, FPDF, or SM (\textit{upper panel}). $\text{RCI}_{\text{ref}=\text{SM}}(\text{Cur.})$ nearly equals zero, indicating little complementary information from the curvature. The lower panel displays the correlation matrix of FPS, FPDF, SM1, SM2, and the curvature statistic, where strong correlations are observed among the curvature, SM1, and SM2.}
    \label{fig: curvature}
\end{figure}
We also employed the curvature statistic,
$\E{|\kappa|}$, introduced by \citet{2011MNRAS.410.1096B}, for posterior-based comparison with \lyanna. However, due to the informational redundancy between the curvature statistic and SM, this statistic was excluded from the comparison analysis. In \Cref{fig: curvature}, the top panel shows the relative complementarity index when the target statistic ($S_t$) is the curvature statistic and the reference statistic ($S_r$) is FPS, FPDF, or SM. The near-zero value $\text{RCI}_{\text{ref}=\text{SM}}(\text{Cur.})$ implies that there is nearly no additional information when SM is combined with the curvature. On the other hand, FPS and FPDF have much higher values of the relative complementarity index, suggesting that the curvature provides additional independent information beyond what is captured by FPS and FPDF. The bottom panel shows the joint correlation matrix of FPS, FPDF, SM1, SM2, and the curvature. The first row of the joint correlation matrix contains the correlation coefficients between the curvature and the rest of the summary vectors, suggesting that SM1 and SM2 are more correlated with curvature than with FPS and FPDF.

\FloatBarrier 
\clearpage

\end{appendix}

\end{document}